\documentclass{article}
 \usepackage{corrinchmargs}
 \usepackage{splat}
 \usepackage{hyperref}

\newcommand{\rigaddr}{Dept.\ of Mathematical and Computer Sciences\\
 Loyola University\\
 6525 N. Sheridan Rd.\\
 Chicago, IL\ \ 60626-5385}

 \input{elsart2article}

\makeatletter  

\newcommand{\comm}[1]{}

\def\uncompactdate#1#2#3#4#5#6#7#8{\ifcase#5#6\or January\or February\or March\or April\or May\or June\or July\or August\or September\or October\or November\or December\fi \space#7#8, #1#2#3#4}

\newcommand{\twodig}[1]{\ifcase#1 00\or01\or02\or03\or04\or05\or06\or07\or08\or09\else#1\fi}

\newcommand{\stretchset}[1]{\renewcommand{\baselinestretch}{#1}\@normalsize}

\renewcommand{\cleardoublepage}{\clearpage\if@twoside \ifodd\c@page\else
    \hbox{}\thispagestyle{plain}\newpage\if@twocolumn\hbox{}\newpage\fi\fi\fi}

\newcommand{\modestsections}{
    \let\part=\section
    \let\section=\subsection
        \renewcommand{\thesubsection}{\arabic{subsection}}
    \let\subsection=\subsubsection  
    \let\subsubsection=\paragraph
    \let\paragraph=\subparagraph
    \renewcommand{\subparagraph}[1]{\paragraph{##1}
        \typeout{You've used the subparagraph command, which is the same as
                 the paragraph command since you're using modest sections.}}
    \renewcommand\appendix{\par
        \setcounter{subsection}{0}%
        \setcounter{subsubsection}{0}%
        \renewcommand\thesubsection{\@Alph\c@subsection}}
}

\newcommand{\modestmaketitle}{
\def\@maketitle{\newpage
 \null
 \vskip 1.5em
 \begin{center}%
  {\Large \@title \par}%
  \vskip 1em
  {\large
   \lineskip .5em
   \begin{tabular}[t]{c}\@author
   \end{tabular}\par}%
  \vskip .7em
  {\large \@date}%
 \end{center}%
 \par
 \vskip 1em}
}

\newcommand{\arabicifpos}[1]{\@arabicifpos{\@nameuse{c@#1}}}
\newcommand{\@arabicifpos}[1]{\ifnum #1>0 \number #1\fi}

\renewcommand{\thefootnote}{\arabicifpos{footnote}}

\def\rect{\@ifstar{\@blackrect}{\@rect}}

\def\square{\@ifstar{\@blacksquare}{\@square}}

\def\@square#1{\@rect(#1,#1)}

\def\@blacksquare#1{\@blackrect(#1,#1)}

\def\@rect(#1,#2){\makebox(0,0)%
{\begin{picture}(#1,#2)\put(0,0){\framebox(#1,#2){}}\end{picture}}}

\def\@blackrect(#1,#2){\makebox(0,0){\rule{#1\unitlength}{#2\unitlength}}}

\def\blackbox(#1,#2)#3{\makebox(#1,#2){\rule{#1\unitlength}{#2\unitlength}}}

\let\blackcircle=\@dot

\newsavebox{\object}
\newsavebox{\arrayputsrow}
\long\def\arrayputs(#1,#2)(#3,#4)#5#6#7{
    \savebox{\object}{#7}
    \savebox{\arrayputsrow}{\multiput(#1,0)(#3,0){#5}{\usebox{\object}}}
    \sbox{\object}{}
    \multiput(0,#2)(0,#4){#6}{\usebox{\arrayputsrow}}
    \sbox{\arrayputsrow}{}}

\def\horizonly@frameb@x#1{%
  \@tempdima\fboxrule
  \advance\@tempdima\fboxsep
  \advance\@tempdima\dp\@tempboxa
  \hbox{%
    \lower\@tempdima\hbox{%
      \vbox{%
        \hrule\@height\fboxrule
        \hbox{%
          #1%
          \vbox{%
            \vskip\fboxsep
            \box\@tempboxa
            \vskip\fboxsep}%
          #1}%
        \hrule\@height\fboxrule}%
                          }%
        }%
}

\newcommand{\horizonlyfboxes}{\let\@frameb@x=\horizonly@frameb@x}

{\begin{minipage}{#1}\begin{tabbing}}%
{\end{tabbing}\end{minipage}}

\newenvironment{tabpackage}
{\begin{minipage}{\columnwidth}\begin{tabbing}}%
{\end{tabbing}\end{minipage}}

\newcommand{\inputverb}[1]%
{\typeout{(#1 (verbatim)}\verbatiminput{#1}\typeout{)}}

\newcount\n
\def\repetition#1#2{\n=0\loop\ifnum\n<#2\advance\n by 1 #1\repeat}

\def\fewreps#1#2{\ifcase#2\or 
#1\or
#1#1\or
#1#1#1\or
#1#1#1#1\or
#1#1#1#1#1\or
#1#1#1#1#1#1\or
#1#1#1#1#1#1#1\or
#1#1#1#1#1#1#1#1\or
#1#1#1#1#1#1#1#1#1\or
#1#1#1#1#1#1#1#1#1#1\or
#1#1#1#1#1#1#1#1#1#1#1\or
#1#1#1#1#1#1#1#1#1#1#1#1\or
#1#1#1#1#1#1#1#1#1#1#1#1#1\or
#1#1#1#1#1#1#1#1#1#1#1#1#1#1\or
#1#1#1#1#1#1#1#1#1#1#1#1#1#1#1\or
#1#1#1#1#1#1#1#1#1#1#1#1#1#1#1#1\or
#1#1#1#1#1#1#1#1#1#1#1#1#1#1#1#1#1\or
#1#1#1#1#1#1#1#1#1#1#1#1#1#1#1#1#1#1\or
#1#1#1#1#1#1#1#1#1#1#1#1#1#1#1#1#1#1#1\or
#1#1#1#1#1#1#1#1#1#1#1#1#1#1#1#1#1#1#1#1\or
#1#1#1#1#1#1#1#1#1#1#1#1#1#1#1#1#1#1#1#1#1\or
#1#1#1#1#1#1#1#1#1#1#1#1#1#1#1#1#1#1#1#1#1#1\or
#1#1#1#1#1#1#1#1#1#1#1#1#1#1#1#1#1#1#1#1#1#1#1\or
#1#1#1#1#1#1#1#1#1#1#1#1#1#1#1#1#1#1#1#1#1#1#1#1\or
#1#1#1#1#1#1#1#1#1#1#1#1#1#1#1#1#1#1#1#1#1#1#1#1#1\or
#1#1#1#1#1#1#1#1#1#1#1#1#1#1#1#1#1#1#1#1#1#1#1#1#1#1\or
#1#1#1#1#1#1#1#1#1#1#1#1#1#1#1#1#1#1#1#1#1#1#1#1#1#1#1\or
#1#1#1#1#1#1#1#1#1#1#1#1#1#1#1#1#1#1#1#1#1#1#1#1#1#1#1#1\or
#1#1#1#1#1#1#1#1#1#1#1#1#1#1#1#1#1#1#1#1#1#1#1#1#1#1#1#1#1\or
#1#1#1#1#1#1#1#1#1#1#1#1#1#1#1#1#1#1#1#1#1#1#1#1#1#1#1#1#1#1\or
#1#1#1#1#1#1#1#1#1#1#1#1#1#1#1#1#1#1#1#1#1#1#1#1#1#1#1#1#1#1#1\or
#1#1#1#1#1#1#1#1#1#1#1#1#1#1#1#1#1#1#1#1#1#1#1#1#1#1#1#1#1#1#1#1\or
#1#1#1#1#1#1#1#1#1#1#1#1#1#1#1#1#1#1#1#1#1#1#1#1#1#1#1#1#1#1#1#1#1\or
#1#1#1#1#1#1#1#1#1#1#1#1#1#1#1#1#1#1#1#1#1#1#1#1#1#1#1#1#1#1#1#1#1#1\or
#1#1#1#1#1#1#1#1#1#1#1#1#1#1#1#1#1#1#1#1#1#1#1#1#1#1#1#1#1#1#1#1#1#1#1\or\else
\typeout{fewreps warning: Number of repetitions outside 0--35 range}\fi}

\makeatother

\newcounter{proglevel}
\newcounter{progline}
\newlength{\linenosep}
\newlength{\subprogindent}
\newlength{\progitemhangindent}
\newlength{\addprogitemsep}

\newenvironment{program}[1]%
{\setcounter{proglevel}{0}\setcounter{progline}{0}
 \settowidth{\linenosep}{\quad}
 \settowidth{\subprogindent}{\qquad}\settowidth{\progitemhangindent}{\quad}
 \setlength{\addprogitemsep}{0pt}
\newcommand{\setsubprogtabs}{#1\=\hspace*{\linenosep}
                             \hspace*{\value{proglevel}\subprogindent}\=
                             \hspace*{\progitemhangindent}\=\+\+\+\kill}
 \begin{tabpackage}\setlength{\tabbingsep}{0pt}\setsubprogtabs}%
{\end{tabpackage}}

\newenvironment{subprogram}%
{\stepcounter{proglevel}\pushtabs\<\<\<\-\-\-\setsubprogtabs}%
{\poptabs\addtocounter{proglevel}{-1}}

\newif\ifprogitemarg
\makeatletter
\def\progitem{\@ifnextchar [{\progitemargtrue\progitemwarg}%
{\progitemargfalse\progitemwarg[\proglinenum]}}
\makeatother

\let\prit=\progitem

\newlength{\progitemstrutheight}
\def\progitemwarg[#1]{\<\< #1 \'\>\rule{0pt}{\progitemstrutheight}%
\ifprogitemarg\else%
\addtocounter{progline}{-1}\refstepcounter{progline}\ignorespaces\fi}

\newcommand{\proglinenum}{\refstepcounter{progline}\theprogline}

\newcommand{\ifthenelse}[3]     
    {{\it if} #1 \begin{description}
                                  \item[\hfill {\it then}] #2
                                  \item[\hfill {\it else}] #3
                              \end{description}}

\newcommand{\var}[1]{\mbox{\it #1}}

\newcommand{\Do}{{\bf do\ }}

\newcommand{\For}{{\bf for\ }}
\newcommand{\Endfor}{{\bf endfor\ }}
\newcommand{\To}{{\bf to\ }}

\newcommand{\If}{{\bf if\ }}
\newcommand{\Then}{{\bf then\ }}
\newcommand{\Else}{{\bf else\ }}

\newcommand{\Endif}{{\bf endif\ }}

\makeatletter

\newcommand{\bibintro}[1]{%
\let\@origthebib=\thebibliography
\let\@origlist=\list
\def\thebibliography{\def\list{#1\par\bigskip\@origlist}\@origthebib}%
}

\makeatother

\newcommand{\inlineciteclosed}
  {\ifx\newblock\undefined\newcommand{\newblock}{\hskip .11em plus .33em minus .07em}\fi}

\inlineciteclosed

\newcounter{oscounter}

\newcommand{\ordinal}[1]{#1\ordsuffix{#1}}

\newcommand{\ordsuffix}[1] 
{\ifcase#1
th\or st\or nd\or rd\or th\or th\or th\or th\or th\or th\or
th\or th\or th\or th\else\setcounter{oscounter}{#1}\ordsuffixrecur\fi}

\newcommand{\ordsuffixrecur}
{\addtocounter{oscounter}{-10}\ifcase\value{oscounter}%
th\or st\or nd\or rd\or th\or th\or th\or th\or th\or th\else
\ordsuffixrecur\fi}

\newcommand{\soda}[1]
    {Proceedings of the \ordinal{#1} Annual {\ifnum#1>3ACM-\fi}SIAM
Symposium on Discrete Algorithms}

\newcommand{\spaa}[1]
    {Proceedings of the
     \ifcase#1\or 1989 \else{\ordinal{#1} Annual }\fi
     ACM Symposium on Parallel Algorithms and Architectures}

\begin{document}

\begin{frontmatter}

\date{} 

\title{Computing the Number of Longest Common Subsequences}

\author{Ronald I. Greenberg\\
 \rigaddr\\
 \url{http://www.cs.luc.edu/~rig}
 }

\maketitle

\begin{abstract}
 This note provides very simple, efficient algorithms for computing the
number of distinct longest common subsequences of two input strings
and for computing the number of LCS embeddings.

\end{abstract}


\end{frontmatter}

\section{Background and Terminologies}
 Let $A=a_1a_2\ldots a_m$ and $B=b_1b_2\ldots b_n$ ($m\leq n$) be two
sequences over an alphabet $\Sigma$.  A sequence that can be obtained
by deleting some symbols of another sequence is referred to as a {\em
subsequence} of the original sequence.  A {\em common subsequence} of
$A$ and $B$ is a subsequence of both $A$ and $B$.  A longest common
subsequence (LCS) is a common subsequence of greatest possible length.
A pair of sequences may have many different LCSs.  In addition, a
single LCS may have many different {\em embeddings}, i.e., positions
in the two strings to which the characters of the LCS correspond.

Most investigations of the LCS problem have focused on efficiently
finding one LCS.  A widely familiar $O(mn)$ dynamic programming
approach goes back at least as far as the early
1970s~\cite{NeedlemanW1970,Sankoff1972,WagnerF1974}, and many later
studies have focused on improving the time and/or space required for
the computation.  Methods have also been developed to efficiently
generate a listing of all distinct LCSs or all LCS embeddings in time
proportional to the output size (plus a preprocessing time of $O(mn)$
or less)~\cite{AltschulE1986,Gotoh1990,Rick2000E,Greenberg2002F}.
Here we show that the simplest scheme~\cite{Greenberg2002F} can be
simplified even further if we seek only a count of the number of
distinct LCSs (or of the number of LCS embeddings).  We obtain a
running time of $O(mn)$ and a space bound of $O(m)$.  (While the
number of LCSs (or LCS embeddings) can grow very large as input size
increases~\cite{Greenberg2003Btechr}, the results here are based on
the standard assumption of unit time for any arithmetic operation
without worrying about the possible magnitude of the operands.)

 \label{sec:intro}

\section{Computing the Number of LCSs or LCS embeddings}
 The familiar $O(mn)$ method for computing the length of an LCS is a
``bottom-up'' dynamic programming approach based on the following
recurrence for the length $L[i,j]$ of an LCS of $a_1a_2\ldots a_i$ and
$b_1b_2\ldots b_j$:
 \begin{equation}
 \label{eqn:naiverecur}
 L[i,j] =
    \left\{
    \begin{tabular}{ll}
    0 & if $i=0$ or $j=0$ \\
    $L[i-1,j-1]+1$ & if $i,j>0$ and $a_i=b_j$ \\
    $\max\{L[i-1,j],L[i,j-1]\}$ & otherwise
    \end{tabular}
    \right.
 \end{equation}

We can use a similar approach to devise an $O(mn)$ algorithm to
compute the number of distinct LCSs $D[m,n]$ of $a_1a_2\ldots a_m$ and
$b_1b_2\ldots b_n$:
 \begin{center}
 \begin{program}{99}
 \prit \For $j\gets0$ \To $n$ \Do \\
 \subprogram
    \prit \For $i\gets0$ \To $m$ \Do \\
    \subprogram
       \prit \If $i=0$ or $j=0$ \Then $D[i,j]\gets1$ \\
       \prit \Else \\
       \subprogram
          \prit $D[i,j]\gets0$\label{line:could-go-in-else} \\
          \prit \If $a_i=b_j$ \Then $D[i,j]\gets D[i-1,j-1]$\label{line:key-then} \\
          \prit \Else\label{line:key-else} \\
          \subprogram
             \prit \If $L[i-1,j]=L[i,j]$ \Then $D[i,j]\gets D[i,j]+D[i-1,j]$ \Endif \\
             \prit \If $L[i,j-1]=L[i,j]$ \Then $D[i,j]\gets D[i,j]+D[i,j-1]$ \Endif \\
             \prit \If $L[i-1,j-1]=L[i,j]$ \Then $D[i,j]\gets D[i,j]-D[i-1,j-1]$ \Endif
                   \label{line:subtract-diag} \\
          \endsubprogram
          \prit \Endif\label{line:key-endif} \\
       \endsubprogram
       \prit \Endif \\
    \endsubprogram
    \prit \Endfor \\
 \endsubprogram
 \prit \Endfor \\
 \end{program}
 \end{center}
 (Note that there is always at least one LCS, since the empty string
$\epsilon$ is always considered to be a common subsequence of the input
sequences.)

In the pseudocode above, line~\ref{line:could-go-in-else} could have
been moved inside the \Else clause beginning at
line~\ref{line:key-else}, but the pseudocode as written is
particularly easy to modify for computation of the number of LCS
embeddings rather than the number of distinct LCSs; just replace that
\Else with the \Endif from line~\ref{line:key-endif}.  (The test in
line~\ref{line:subtract-diag} is never satisfied with $a_i=b_j$, but
it is harmless and concise to write the code this way.)

We may also note that $O(m)$ space suffices for the computation, since
we really only need a portion of two columns of the $L$ and $D$ arrays
at any time.  Here is a rewrite of the code to achieve $O(m)$ space
that also introduces the necessary change to switch from computing the
number of distinct LCSs to computing the number of LCS embeddings
$E[m]$; we also include the efficient computation of the $L$ values:
 \begin{center}
 \begin{program}{99}
 \prit \For $j\gets0$ \To $n$ \Do \\
 \subprogram
    \prit \For $i\gets0$ \To $m$ \Do \\
    \subprogram
       \prit \If $i=0$ or $j=0$ \Then $L[i]\gets0$, $\var{oldL}\gets0$,
                                      $E[i]\gets1$, and $oldE\gets1$ \\
       \prit \Else \\
       \subprogram
          \prit $\var{newL}\gets\max\{L[i-1],L[i]\}$ and $\var{newE}\gets0$ \\
          \prit \If $a_i=b_j$ \Then $\var{newL}\gets\var{oldL}+1$ and $\var{newE}\gets\var{oldE}$ \Endif \\
             \prit \If $L[i-1]=\var{newL}$ \Then $\var{newE}\gets \var{newE}+E[i-1]$ \Endif \\
             \prit \If $L[i]=\var{newL}$ \Then $\var{newE}\gets \var{newE}+E[i]$ \Endif \\
             \prit \If $\var{oldL}=\var{newL}$ \Then $\var{newE}\gets \var{newE}-\var{oldE}$ \Endif \\
          \prit $\var{oldL}\gets L[i]$, $\var{oldE}\gets E[i]$,
                $L[i]\gets\var{newL}$, and $E[i]\gets\var{newE}$ \\
       \endsubprogram
       \prit \Endif \\
    \endsubprogram
    \prit \Endfor \\
 \endsubprogram
 \prit \Endfor
 \end{program}
 \end{center}

 \label{sec:compute}

\bibliographystyle{plain}
\bibliography{sources}

\begin{thebibliography}{1}

\bibitem{AltschulE1986}
Stephen~F. Altschul and Bruce~W. Erickson.
\newblock Optimal sequence alignment using affine gap costs.
\newblock {\em Bulletin of Mathematical Biology}, 48(5/6):603--616, 1986.

\bibitem{Gotoh1990}
Osamu Gotoh.
\newblock Optimal sequence alignment allowing for long gaps.
\newblock {\em Bulletin of Mathematical Biology}, 52(3):359--373, 1990.

\bibitem{Greenberg2002F}
Ronald~I. Greenberg.
\newblock Fast and simple computation of all longest common subsequences.
\newblock Eprint arXiv:cs.DS/0211001, Comp.\ Sci.\ Res.\ Repository,
  \url{http://arXiv.org/abs/cs.DS/0211001}, 2002.

\bibitem{Greenberg2003Btechr}
Ronald~I. Greenberg.
\newblock Bounds on the number of longest common subsequences.
\newblock Technical Report arXiv:cs.DM/0301030, Comp.\ Sci.\ Res.\ Repository,
  \url{http://arXiv.org/abs/cs.DM/0301030}, 2003.

\bibitem{NeedlemanW1970}
Saul~B. Needleman and Christian~D. Wunsch.
\newblock A general method applicable to the search for similarities in the
  amino acid sequence of two proteins.
\newblock {\em Journal of Molecular Biology}, 48:443--453, 1970.

\bibitem{Rick2000E}
Claus Rick.
\newblock Efficient computation of all longest common subsequences.
\newblock In {\em \swat{7}}, volume 1851 of {\em Lecture Notes in Computer
  Science}, pages 407--418. Springer-Verlag, 2000.

\bibitem{Sankoff1972}
David Sankoff.
\newblock Matching sequences under deletion/insertion constraints.
\newblock {\em Proceedings of the National Academy of Science USA}, 69(1):4--6,
  January 1972.

\bibitem{WagnerF1974}
Robert~A. Wagner and Michael~J. Fischer.
\newblock The string-to-string correction problem.
\newblock {\em Journal of the ACM}, 21(1):168--173, 1974.

\end{thebibliography}

\end{document}